\newcommand{\beeq}{\begin{equation}}
\newcommand{\eneq}{\end{equation}}
\newcommand{\be}{\begin{eqnarray}}
\newcommand{\ee}{\end{eqnarray}}
\newcommand{\bpic}{\begin{picture}}
\newcommand{\epic}{\end{picture}}
\newcommand{\bs}{\begin{scriptsize}}
\newcommand{\es}{\end{scriptsize}}

\def\dd{\partial}
\def\la{\raise.16ex\hbox{$\langle$} \, }
\def\ra{\, \raise.16ex\hbox{$\rangle$} }

\def\a{\alpha}
\def\b{\beta}
\def\d{\delta}
\def\e{\epsilon}
\def\g{\gamma}
\def\G{\Gamma}
\def\r{\rho}
\def\s{\sigma}
\def\O{\Omega}
\def\D{\Delta}
\def\L{\Lambda}
\def\l{\lambda}

\def\Box{\kern1pt\vbox{\hrule height 1.2pt\hbox{\vrule width 1.2pt\hskip 3pt
   \vbox{\vskip 6pt}\hskip 3pt\vrule width 0.6pt}\hrule height 0.6pt}\kern1pt}
\def\gtwid{\mathrel{\raise.3ex\hbox{$>$\kern-.75em\lower1ex\hbox{$\sim$}}}}
\def\ltwid{\mathrel{\raise.3ex\hbox{$<$\kern-.75em\lower1ex\hbox{$\sim$}}}}
\documentstyle[epsfig,12pt]{article}

\begin{document}
\begin{titlepage}
\begin{flushright}
gr-qc/0204065 \\ UFIFT-HEP-02-13
\end{flushright}
\vspace{.4cm}
\begin{center}
\textbf{Super-Acceleration from Massless, Minimally Coupled $\phi^4$}
\end{center}
\begin{center}
V. K. Onemli$^{\dagger}$ \\
R. P. Woodard$^{\ddagger}$
\end{center}
\begin{center}
\textit{Department of Physics \\ University of Florida \\
Gainesville, FL 32611 USA}
\end{center}
\begin{center}
ABSTRACT
\end{center}
\hspace*{.5cm} We derive a simple form for the propagator of a massless,
minimally coupled scalar in a locally de Sitter geometry of arbitrary spacetime
dimension. We then employ it to compute the fully renormalized stress tensor
at one and two loop orders for a massless, minimally coupled $\phi^4$ theory
which is released in Bunch-Davies vacuum at $t=0$ in co-moving coordinates. 
In this system the uncertainty principle elevates the scalar above the minimum 
of its potential, resulting in a phase of super-acceleration. With the 
non-derivative self-interaction the scalar's breaking of de Sitter invariance 
becomes observable. It is also worth noting that the weak energy condition is 
violated on cosmological scales. An interesting subsidiary result is that 
canceling overlapping divergences in the stress tensor requires a conformal 
counterterm which has no effect on purely scalar diagrams.
\begin{flushleft}
\end{flushleft}
\vspace{.4cm}
\begin{flushleft}
$^{\dagger}$ e-mail: onemli@phys.ufl.edu \\
$^{\ddagger}$ e-mail: woodard@phys.ufl.edu
\end{flushleft}
\end{titlepage}

\section{Introduction}

One of the most intriguing and potentially important results of observational
cosmology is the inference, from Type Ia supernovae at high redshift
\cite{Reiss,Perlmutter}, that the universe is entering a phase of ``accelerated 
expansion''. This phrase sometimes gives rise to a misconception that is best
explained in the context of a homogeneous, isotropic and spatially flat 
geometry,
\be
ds^2 = -dt^2 + a^2(t) d\vec{x} \cdot d\vec{x} \; .
\ee
The rate of cosmological expansion is the Hubble constant,
\be
H(t) \equiv {\dot{a}(t) \over a(t)} \; ,
\ee
whereas the deceleration parameter is,
\be
q(t) \equiv -1 - {\dot{H}(t) \over H^2(t)} \; .
\ee
Although the general public sometimes takes ``accelerated expansion'' to mean 
that $\dot{H} > 0$, the actual meaning is rather that $q < 0$. 

In fact no stable theory can exhibit $\dot{H} > 0$ on the classical level. 
This is a simple consequence of the two nontrivial Einstein equations in 
this geometry,
\begin{eqnarray}
3 H^2 & = & 8\pi G \rho \; , \\
-2 \dot{H} - 3 H^2 & = & 8 \pi G p \; . 
\end{eqnarray}
Here $\rho(t)$ and $p(t)$ are the energy density and pressure, respectively.
Adding the two equations gives,
\be
-2 \dot{H} = 8 \pi G (\rho + p) \; .
\ee
The quantity on the right-hand side is non-negative as a consequence of the
Weak Energy Condition \cite{Hawking} which asserts that no physical observer
can see a negative energy density. Its mathematical transcription is
$T_{\mu\nu} W^{\mu} W^{\nu} \geq 0$ for all timelike vectors $W^{\mu}$. To 
derive $\rho + p \geq 0$ for a homogeneous and isotropic cosmology, note first 
that the nonzero elements of the stress-energy tensor are $T_{00} = - \rho 
g_{00}$ and $T_{ij} = p g_{ij}$. Now apply the condition in the limit that 
$W^{\mu}$ becomes null.
 
The Weak Energy Condition not only implies $\dot{H} \leq 0$, it also tells us
that $q \geq -1$. Although stable theories obey the Weak Energy Condition on 
the classical level it has long been known that quantum effects can give rise
to violations \cite{Ford}, although those studied so far have been on 
microscopic scales. This is of great interest because the present data is 
actually consistent with $q < -1$ \cite{Caldwell}. It is only theoretical 
prejudice in favor of stability and against quantum effects on cosmological
scales that results in the usual likelihood contours being cut off at $q = -1$.
If the proposed Supernova Pencil Beam Survey \cite{Wang} or SNAP experiments 
\cite{SNAP} were to give a definitive determination of $q < -1$ we should be 
forced to abandon this prejudice. The purpose of this paper is to give a 
precise formulation of a field theory in which quantum cosmological effects 
do in fact result in $\rho + p < 0$ for an arbitrarily long period of time.

The model has been studied before, as has its potential for violating the Weak 
Energy Condition on cosmological scales \cite{AbWo}. It consists of a 
massless, minimally coupled scalar with a quartic self-interaction. Although
gravity is not dynamical, neither is the background flat. The scalar is a 
spectator to $\Lambda$-driven inflation. The new feature, which is the subject 
of this paper, is that we can now regulate the model in a simple way that 
preserves general coordinate invariance. The acid test of simplicity is that
one can go beyond coincident propagators to explicitly evaluate higher loop 
graphs which involve integrations. We shall demonstrate this by computing all 
one and two loop contributions to the expectation value of the stress-energy 
tensor in the presence of a state which is Bunch-Davies vacuum at $t=0$. Our 
result confirms the conjecture of ref. \cite{AbWo} that this model shows 
super-acceleration, i.e., $\rho + p < 0$. The model should also be of interest 
as an exercise in quantum field theory on curved spacetime in which the 
computations are pushed beyond the one-loop approximation that has been so 
thoroughly studied \cite{BD}.

This Introduction is the first of eight sections. In Section 2 we present the
Lagrangian and describe the various diagrams which comprise the expectation 
value of the stress-energy tensor at one and two loop order. Section 3 gives
an explicit and relatively simple expression for the $D$-dimensional 
propagator, the possession of which is what allows us to apply dimensional 
regularization at arbitrary order. In Section 4 we compute the one loop mass
counterterm and the one loop expectation value of the stress-energy tensor.
Section 5 evaluates the leading order, local diagrams that were already 
obtained in ref. \cite{AbWo}. The power of our dimensional regularization
technique is displayed in Section 6 by computing the nonlocal diagrams which 
could not be done previously. Although these terms do not contribute at 
leading order (as conjectured previously \cite{AbWo}) they are necessary to 
make the stress-energy tensor conserved and they do probe the nonlocal,
ultraviolet finite sector of the theory. They also contribute some nonlocal
ultraviolet divergences which, if uncompensated, would compromise the model's
renormalizability. Section 7 evaluates the expectation value of the conformal
counterterm which absorbs them. (The need for this was also realized before
\cite{AbWo}.) The fully renormalized result is presented in Section 8.

\section{The model and its stress-energy tensor}

Our model has the following Lagrangian,
\be
{\cal L} = -\frac12 \partial_{\mu} \phi \partial_{\nu} \phi g^{\mu\nu}
\sqrt{-g} - \frac1{4!} \lambda \phi^4 \sqrt{-g} + {\Delta {\cal L}} \;.
\label{Lag}
\ee
The counterterms reside in ${\Delta {\cal L}}$,
\be
{\Delta {\cal L}} & = & -\frac12 {\delta m^2} \phi^2 \sqrt{-g} +
{\delta \xi} \Bigl(R - D (D-1) H^2\Bigr) \phi^2 \sqrt{-g} - {{\delta \Lambda} 
\over 8 \pi G} \sqrt{-g} \; , \nonumber \\
& & \qquad \qquad - \frac12 {\delta Z} \partial_{\mu} \phi \partial_{\nu}
\phi g^{\mu\nu} \sqrt{-g} - \frac1{4!} {\delta \lambda} \phi^4 \sqrt{-g} \; .
\ee
Those on the second line are of order $\lambda^2$, just as in flat space
\cite{Peskin}, and are irrelevant for the purposes of this paper. However,
the counterterms on the top line are needed to remove divergences at one and
two loop order in the stress-energy tensor. It turns out that ${\delta m^2}$
is of order $\lambda$ while ${\delta \Lambda}$ is actually of order 
$\lambda^0$. The conformal counterterm, ${\delta \xi}$ is needed to remove
overlapping divergences in the expectation value of $T_{\mu\nu}$ at order
$\lambda^1$. Note that this term makes no contribution at all to pure scalar
interactions on account of the fact that $R = D (D-1) H^2$ in $D$-dimensional,
locally de Sitter background.

Only the scalar is quantized. The metric is a non-dynamical background which 
we take to be locally de Sitter geometry with cosmological constant $\Lambda 
= (D-1) H^2$. It is most convenient to work in conformal coordinates for 
which the metric takes the form,
\begin{equation}
g_{\mu \nu}(\eta,\vec{x}) = \Omega^2(\eta) \eta_{\mu\nu} \qquad \mbox{where}
\qquad \Omega(\eta) = -{1 \over H \eta} \;.
\end{equation}
It is useful to keep in mind the relation to co-moving coordinates,
\be
\eta = - H^{-1} e^{-Ht} \qquad \Leftrightarrow \qquad t = H^{-1} \ln\Bigl(
\Omega(\eta)\Bigr) \; .
\ee
We release the state in Bunch-Davies vacuum at $t = 0$, corresponding to 
conformal time $\eta = -H^{-1}$. Note that the infinite future corresponds 
to $\eta \rightarrow 0^-$, so the possible variation of causally related 
conformal coordinates in either space or time is at most ${\Delta x} = 
{\Delta \eta} = H^{-1}$.

The stress-energy tensor is,
\begin{eqnarray}
\lefteqn{ T_{\mu\nu}(x) \equiv \frac{-2}{\sqrt{-g(x)}} \frac{\d S_{\tiny
{matter}}}{\d g^{\mu\nu}(x)} \; , } \\
& = & (1 + {\d Z}) \left[ \d^{\r}_{~\mu} \d^{\sigma}_{~\nu} - \frac12 
g_{\mu\nu}(x) g^{\r\s}(x) \right] \dd_{\r} \phi(x) \dd_{\s} \phi(x) \nonumber\\
& & - g_{\mu\nu}(x) \frac{\l + \d \l}{4!} \phi^4(x) - \frac12 {\d m^2} 
g_{\mu\nu}(x) \phi^2(x) - \frac{\d\L}{8\pi G}g_{\mu\nu} \nonumber \\
& & - 2 {\d \xi} \left[(D-1) H^2 \phi^2 g_{\mu\nu} + g_{\mu\nu} (\phi^2)^{;
\rho}\,_\rho - (\phi^2)_{;\mu\nu} \right] \; . \label{T} 
\end{eqnarray}
We want the one and two loop contributions to the expectation value of this 
operator in the state which is Bunch-Davies vacuum at $t=0$. If $i\Delta(x;x')$
stands for the bare propagator in this state, the lowest order kinetic energy
contributions are,
\begin{eqnarray}
\lefteqn{\left[ \d^{\r}_{~\mu} \d^{\s}_{~\nu} - \frac12 g_{\mu\nu}(x) 
g^{\r\s}(x) \right] \Bigl\langle \O \Bigl\vert \dd_{\r} \phi(x) \dd_{\s} 
\phi(x) \Bigr\vert \O \Bigr\rangle } \nonumber \\
& & = \left[ \d^{\r}_{~\mu} \d^{\s}_{~\nu} - \frac12 g_{\mu\nu}(x) 
g^{\r\s}(x) \right] \Bigl\{\dd_{\r} \dd_{\s}^{\prime} i\Delta(x;x') 
\Bigl\vert_{x' \rightarrow x} + O(\lambda)\Bigr\} \; . \label{kinetic}
\end{eqnarray}
The first term is evaluated in Section 4 and the order $\lambda$ correction
is computed in Section 6. The lowest order contributions from the potential
energy are,
\begin{eqnarray}
\lefteqn{-g_{\mu\nu}(x) \Bigl\langle \O \Bigl\vert \frac{\lambda}{4!} \phi^4(x)
+ \frac{\delta m^2}2 \phi^2(x) \Bigr\vert \O \Bigr\rangle} \nonumber \\
& & = -g_{\mu\nu}(x) \left\{ \frac{\lambda}8 \left[i\Delta(x;x)\right]^2 +
\frac{\delta m^2}2 i\Delta(x;x) + O(\lambda^2)\right\} \; . \label{pot}
\end{eqnarray}
This is evaluated in Section 5. The contribution from the conformal 
counterterm is,
\begin{eqnarray}
\lefteqn{- 2\delta\xi \Bigl\{ (D-1) H^2 g_{\mu\nu}(x) \left\langle \O 
\left\vert \phi^2(x) \right\vert \O \right\rangle } \nonumber \\
& & \hspace{.5cm} + \left[g_{\mu\nu}(x) g^{\r\s}(x) - \d^{\r}_{~\mu} 
\d^{\s}_{~\nu} \right] \left(\dd_{\r} \dd_{\s} - \Gamma^{\alpha}_{~\r\s}(x)
\dd_{\alpha} \right) \left\langle \O \left\vert \phi^2(x) \right\vert \O 
\right\rangle \Bigr\} \nonumber \\
& & = - 2\delta\xi \left\{ (D-1) H^2 g_{\mu\nu}(x) i\Delta(x;x) \right. 
\nonumber \\
& & \hspace{.5cm} \left. + \left[g_{\mu\nu}(x) g^{\r\s}(x) - \d^{\r}_{~\mu} 
\d^{\s}_{~\nu} \right] \left( \dd_{\r} \dd_{\s} - \Gamma^{\alpha}_{~\r\s}(x)
\dd_{\alpha} \right) i\Delta(x;x) + O(\lambda^2) \right\} . \qquad \label{conf}
\end{eqnarray}
This term is evaluated in Section 7.

Obtaining a true expectation value, rather than an in-out matrix element, 
requires use of the Schwinger-Keldysh formalism \cite{Schwinger,Jordan}. 
However, this formalism reduces to the usual Feynman diagrams for the lowest 
order graphs. We shall not actually need the full Schwinger-Keldysh formalism 
until discussing order $\lambda$ kinetic energy corrections in Section 5.

Because the state is homogeneous and isotropic, the nonzero tensor components 
are $\langle \O \vert T_{00} \vert \O \rangle = -\rho g_{00}$ and $\langle \O
\vert T_{ij} \vert \O \rangle = p g_{ij}$. This is true as well for each of 
the three contributions just described, so we shall finally report an energy
density and pressure for each. Although neither the kinetic energy nor the 
potential energy contributions is separately conserved, their sum is. The 
conformal contribution is conserved by itself, as is the cosmological
counterterm. However, because these quantities harbor divergences, 
``conservation'' must be understood with the ultraviolet regulator still on. 
That is, we must keep the spacetime dimension $D$ arbitrary, which 
makes conservation read, $\dot{\rho} = - (D-1) H (\rho + p)$. This is an 
important check on accuracy. Another important check is the cancellation of
overlapping divergences that occurs when the various contributions are 
combined.

\section{The $D$-dimensional propagator} \label{section:DP}

The behavior of free massless minimally coupled scalar field has been 
investigated extensively \cite{BAAF,LFLP,FSW,AVLF,L,S}. Among the curious 
properties of these particles are the absence of normalizable, de Sitter
invariant states \cite{BAAF} and the appearance of acausal infrared 
singularities when the Bunch-Davies vacuum is used with infinite spatial
surfaces \cite{LFLP,FSW}. To regulate this infrared problem we work on the 
manifold $T^{D-1} \times R$, with the spatial coordinates in the finite range, 
$-H^{-1}/2 < x^i \leq H^{-1}/2$. Although the actual propagator is a mode sum
on this manifold, the small possible variation in conformal coordinates 
renders the first term of the Euler-Maclaurin formula --- just the integral 
--- an excellent approximation, and the finite spatial range serves merely to
cut off what would have been a logarithmic infrared divergence on infinite
space. In $D = 3 + 1$ spacetime dimensions the result is \cite{TsWo1},
\begin{equation}
{i \Delta}(x;x')\Bigl\vert_{D=4} = \left({H \over 2 \pi} \right)^2 \left\{ 
\frac1{y(x;x')} - \frac12 \ln\Bigl(y(x;x')\Bigr) + \frac12 \ln\Bigl(
\Omega(\eta) \Omega(\eta') \Bigr) \right\} \; , \label{D=4prop}
\end{equation}
where the modified de Sitter length function is,\footnote{What is termed 
``the de Sitter length function'' in the literature is,
\begin{eqnarray}
z(x;x') = 1 - y(x;x') \; . \nonumber
\end{eqnarray}
The geodesic length from $x^{\mu}$ to $x^{\prime \mu}$, $\ell(x;x')$, is
related to $y(x;x')$ as follows,
\begin{eqnarray}
y(x;x') = \sin^2\left(\frac12 H \ell(x;x') \right) \; . \nonumber
\end{eqnarray}}
\begin{equation}
y(x;x') \equiv \Omega(\eta) \Omega(\eta') H^2 \left[ \Vert \vec{x} -
\vec{x}' \Vert^2 - (\vert \eta - \eta' \vert - i e)^2 \right] \; .
\end{equation}

Neglecting the higher order Euler-Maclaurin terms does not prevent 
(\ref{D=4prop}) from solving the correct differential equation. The higher
terms also drop out of quite complicated, nonlinear relations such as the 
Ward Identity for the one loop graviton self-energy \cite{TsWo2}. We shall
therefore regard the technique as valid and confine ourselves to finding the 
appropriate generalization of (\ref{D=4prop}) to $D$ spacetime dimensions. 

We seek a function of $y(x;x')$ and the two conformal factors which obeys the 
equation,
\be
\eta^{\mu\nu} {\partial \over \partial x^{\mu}} \Omega^{D-2}(\eta) {\partial 
\over \partial x^{\nu}} {i \Delta}(x;x') = i \delta^D(x-x') \; . \label{kin}
\ee
One consequence of de Sitter invariance is that the kinetic operator takes
a particularly simple form when acting on a function of just $y(x;x')$. To
reach this form note first that derivatives of $y(x;x')$ give,
\begin{eqnarray}
{\partial y \over \partial x^{\mu}} \!\!\! & = & \!\!\! H \Omega(\eta) \Bigl[y 
\delta^0_{~\mu} + 2 \Omega(\eta') H {\Delta x}_{\mu} + 2 i e H \Omega(\eta') 
{\rm sgn}(\eta - \eta') \delta^0_{~\mu} \Bigr] \; , \\
{\partial^2 y \over \partial x^{\mu} \partial x^{\nu}} \!\!\! & = & \!\!\! H^2 
\Omega^2(\eta) \Bigl[2 y \delta^0_{~\mu} \delta^0_{~\nu} \! + \! 2 
\Omega(\eta') H {\Delta x}_{\mu} \delta^0_{~\nu} \! + \! 2 \Omega(\eta') 
\delta^0_{~\mu} H {\Delta x}_{\nu} \! + \! 2 {\scriptstyle \frac{\Omega(\eta')
}{\Omega(\eta)}} \eta_{\mu\nu} \nonumber \\
& & \quad + 4 i e H \Omega(\eta') {\rm sgn}(\eta - \eta') \delta^0_{~\mu}
\delta^0_{~\nu} + 4 i e \delta(\eta - \eta') \delta^0_{~\mu} \delta^0_{~\nu}
\Bigr] \; .
\end{eqnarray}
Now use $\Omega(\eta') H {\Delta x}^0 = 1 - \Omega(\eta')/\Omega(\eta)$ to 
conclude,
\begin{eqnarray}
\eta^{\mu\nu} \delta^0_{~\mu} {\partial y \over \partial x^{\nu}} & = & H 
\Omega(\eta) \Bigl[-y + 2 - 2 {\scriptstyle \frac{\Omega(\eta')}{\Omega(\eta)}}
- 2 i e H \Omega(\eta') {\rm sgn}(\eta - \eta') \Bigr] \; , \label{con1} \\
\eta^{\mu\nu} {\partial y \over \partial x^{\mu}} {\partial y \over \partial
x^{\nu}} & = & H^2 \Omega^2(\eta) \Bigl[-y^2 + 4 y - 4 i e H \Omega(\eta') 
y {\rm sgn}(\eta - \eta') \Bigr] \; , \label{con2} \\
\eta^{\mu\nu} {\partial^2 y \over \partial x^{\mu} \partial x^{\nu}} & = & H^2 
\Omega^2(\eta) \Bigl[-2 y + 4 + 2 (D - 2) {\scriptstyle \frac{\Omega(\eta')}{
\Omega(\eta)}} \nonumber \\
& & \quad - 4 i e H \Omega(\eta') {\rm sgn}(\eta - \eta') - 4 i e 
\delta(\eta - \eta') \Bigr] \; . \label{con3}
\end{eqnarray}
Finally, use the chain rule and substitute (\ref{con1}-\ref{con3}),
\begin{eqnarray}
\lefteqn{\eta^{\mu\nu} {\partial \over \partial x^{\mu}} \Omega^{D-2}(\eta) 
{\partial \over \partial x^{\nu}} f\Bigl(y(x;x')\Bigr) } \nonumber \\
& & = \Omega^{D-2}(\eta) \eta^{\mu\nu} \Biggl\{ {\partial y \over \partial 
x^{\mu}} {\partial y \over \partial x^{\nu}} f''(y) + {\partial^2 y \over
\partial x^{\mu} \partial x^{\nu}} f'(y) \nonumber \\
& & \hspace{3cm} + (D - 2) H \Omega(\eta) \delta^0_{~\mu} {\partial y \over 
\partial x^{\nu}} f'(y) \Biggr\} \; , \\
& & = H^2 \Omega^D(\eta) \Biggl\{ (4 y - y^2) f^{\prime\prime}(y) + D (2 - y) 
f'(y) - 4 i e \delta(\eta - \eta') f'(y) \nonumber \\
& & \hspace{3cm} -2 i e H \Omega(\eta') {\rm sgn}(\eta - \eta') \Bigl[2 y 
f''(y) + D f'(y)\Bigr] \Biggr\} \; .
\end{eqnarray}

Before considering possible dependence upon the scale factor let us note
that the delta function on the right hand side of (\ref{kin}) descends from a 
factor of $y^{1 - \frac{D}2}$ in the limit that $e \rightarrow 0$. To see 
this, suppose $f(y)$ has the form,
\begin{equation}
f(y) = {k_1 \over y^{\frac{D}2 - 1}} + O\Bigl(y^{-\frac{D}2}\Bigr) \; ,
\end{equation}
where $k_1$ is a constant. Acting the kinetic operator gives,
\begin{equation}
\partial^{\mu} \Omega^{D-2} \partial_{\mu} f(y) = H^2 \Omega^D \left\{ -4 i e
\delta(\eta - \eta') {-k_1 (\frac{D}2 - 1) \over y^{\frac{D}2}} + O\Bigl(y^{1
- \frac{D}2}\Bigr) \right\} \; .
\end{equation}
Now multiply this by a test function (of $x^{\mu}$), integrate $\int d^Dx$, and 
take $e \rightarrow 0$. For $e = 0$, terms of order $y^{1-\frac{D}2}$ diverge 
at $x^{\mu} = x^{\prime \mu}$, but the singularity is integrable. The only 
delta function comes from the term proportional to,
\begin{equation}
e \delta(\eta - \eta') {1 \over y^{\frac{D}2}(x;x')} = {\delta(\eta - \eta') 
\over H^D \Omega^D} {e \over (\Vert \vec{x} - \vec{x}' \Vert^2 + e^2)^{
\frac{D}2}} \longrightarrow {\pi^{\frac{D}2} \over \Gamma(\frac{D}2)} 
{\delta^D(x - x') \over H^D \Omega^D} \; .
\end{equation}
Comparison with (\ref{kin}) fixes $k_1$,
\begin{equation}
k_1 = \left({H \over 2 \pi}\right)^2 \left({\pi \over H^2}\right)^{2-
\frac{D}2} \Gamma({\scriptstyle \frac{D}2} - 1) \; .
\end{equation}

Now consider adding to $f(y)$ a symmetric function of $\Omega(\eta)$ and
$\Omega(\eta')$. The only function which can give the same prefactor of 
$\O^D(\eta)$ is a constant times the same one that appears in (\ref{D=4prop}). 
The $D$-dimensional propagator must therefore take the form,
\be
{i\Delta(x;x')} = f\Bigl( y(x;x')\Bigr) + k_2 \ln\Bigl( \O(\eta) \O(\eta')
\Bigr) \; ,
\ee
In the $e \rightarrow 0$ limit the function $f(y)$ must obey,
\begin{eqnarray}
\lefteqn{H^2 \O^D(\eta) \Bigl\{ (4 y - y^2) f^{\prime\prime}(y) + D (2 - y) 
f'(y) } \nonumber \\
& & - 4 i e \delta(\eta - \eta') f'(y) - k_2 (D-1) \Bigr\} = i \delta^D(x - x') 
\; .  \label{propeq}
\end{eqnarray}

We have seen that getting the correct delta function requires a term of the
form $y^{1 - \frac{D}2}$ plus less singular powers. Series solution generates 
the higher powers. Defining $D \equiv 4 - \epsilon$ gives,
\be
{H^{2 - \epsilon} \over 4 \pi^{2 - \frac{\epsilon}2} } 
\Gamma\left(1 - \frac{\epsilon}2\right) \left\{ {1 \over y^{1 - \frac{
\epsilon}2}} - \left(1 - \frac{\epsilon}2 \right) \sum_{n = 0}^{\infty}
{1 \over n + \frac{\epsilon}2} {\Gamma(3 + n - \frac{\epsilon}2) \over 
(n+1)! \Gamma(2 - \frac{\epsilon}2)} {y^{n + \frac{\epsilon}2} \over 4^{n+1}} 
\right\} \; .
\ee
This series solves (\ref{propeq}), for $k_2=0$, but it does not reduce to 
(\ref{D=4prop}) for $\epsilon = 0$. The $n=0$ term is not even finite in 
this limit! The resolution to both problems is a series of strictly nonnegative
integer powers of $y$, which cancels the divergence and the unwanted terms.
This series obeys the homogeneous equation up to a $y^0$ term which is canceled
by the $k_2 (D-1)$ term. The final result is,
\begin{eqnarray}
\lefteqn{{i \Delta}(x;x') = \left({H \over 2 \pi}\right)^2 \left({H \over
\sqrt{\pi}} \right)^{-\epsilon} \Gamma\left(1 - \frac{\epsilon}2\right)
\left\{ {1 \over y^{1 - \frac{\epsilon}2}} + \left(1 - \frac{\epsilon}2 \right)
\left(1 - \frac{\epsilon}4 \right) \left({1 - y^{\frac{\epsilon}2} \over
\epsilon} \right) \right. } \nonumber \\
& & \hspace{1cm} + \left(1 - \frac{\epsilon}2 \right) \sum_{n = 1}^{\infty}
\left[\frac1{n} {\Gamma(3 + n - \epsilon) \over \Gamma(2 + n - \frac{\epsilon
}2)} - {1 \over n + \frac{\epsilon}2} {\Gamma(3 + n - \frac{\epsilon}2) \over 
(n+1)! \Gamma(2 - \frac{\epsilon}2)} y^{\frac{\epsilon}2} \right] {y^n \over
4^{n+1}} \nonumber \\
& & \hspace{6.5cm} \left. + \frac14 {\Gamma(3 - \epsilon) \over \Gamma(1 -
\frac{\epsilon}2)} \ln\Bigl( \Omega(\eta) \Omega(\eta') \Bigr) \right\} \; .
\label{prop}
\end{eqnarray}

The great advantage of this regularization is that it preserves general
coordinate invariance (once $e$ is taken to zero). One might think that the 
propagator is unwieldy but this is not so in practice. The really cumbersome 
part is the infinite sum on the second line. But these terms all vanish at
coincidence ($y(x;x) = 0$) and they vanish for all $y(x;x')$ at $D=4$. So
one need only retain the higher terms when they multiply something else that
diverges like $1/\epsilon$. Note also that one need never worry about large 
$y(x;x')$ on account of causality.

All valid regularizations must reproduce the result of Vilenkin and Ford
that the coincidence limit of the propagator contains a finite term which
grows like $\ln(\Omega) = H t$ \cite{AVLF,L,S}. To check this note that $y(x;x) 
=0$ at coincidence. When a variable vanishes like this in dimensional 
regularization one must always assume $\epsilon$ to be large enough that
the variable is raised to only nonnegative powers. We therefore find,
\begin{equation}
\lim_{x' \rightarrow x} {i\Delta}(x;x') = \left({H \over 2 \pi}\right)^2
\left({H \over \sqrt{\pi}}\right)^{-\epsilon} \left\{ \frac1{2 \epsilon}
\Gamma\left(3 - \frac{\epsilon}2\right) + \frac12 \Gamma\Bigl(3 - \epsilon
\Bigr) \ln\Bigl(\Omega(\eta)\Bigr) \right\} \; . \label{coinc}
\end{equation}
Note that (\ref{coinc}) is exact for {\it arbitrary} $\epsilon$.

For every sort of line in the standard Feynman rules the Schwinger-Keldysh 
formalism has a ``$+$'' line and a ``$-$'' line of the same sort 
\cite{Schwinger,Jordan}. Although vertices involve either all $+$ or all $-$
lines, there are $++$, $+-$, $-+$ and $--$ propagators. All of them are the
same function (\ref{prop}) of the appropriate version of the modified de
Sitter length function,
\begin{eqnarray}
y_{++}(x;x') & \equiv & \Omega(\eta) \Omega(\eta') H^2 \left[ \Vert \vec{x} -
\vec{x}' \Vert^2 - (\vert \eta - \eta' \vert - i e)^2 \right] \; , \\
y_{+-}(x;x') & \equiv & \Omega(\eta) \Omega(\eta') H^2 \left[ \Vert \vec{x} -
\vec{x}' \Vert^2 - (\eta - \eta' + i e)^2 \right] \; , \\
y_{-+}(x;x') & \equiv & \Omega(\eta) \Omega(\eta') H^2 \left[ \Vert \vec{x} -
\vec{x}' \Vert^2 - (\eta - \eta' - i e)^2 \right] \; , \\
y_{--}(x;x') & \equiv & \Omega(\eta) \Omega(\eta') H^2 \left[ \Vert \vec{x} -
\vec{x}' \Vert^2 - (\vert \eta - \eta' \vert + i e)^2 \right] \; .
\end{eqnarray}
When the points are spacelike related $y(x;x')$ is positive; when they are 
timelike $y(x;x')$ is negative. So ${i\Delta}_{++}(x;x')$ and ${i\Delta}_{
+-}(x;x')$ are equal (in the limit that $e$ vanishes) for all spacelike 
separated points and for $\eta' > \eta$. That is why the $++$ and $+-$ 
contributions cancel whenever $x^{\mu\prime}$ strays outside the past 
lightcone of $x^{\mu}$. Note that inside the past lightcone the $++$ and $+-$
propagators are conjugate.

\section{The one loop counterterms} \label{section:MC}

The one loop ($\lambda^0$) kinetic energy contribution from (\ref{kinetic})
provides a fine illustration of how (\ref{prop}) is used to compute
dimensionally regulated results. Since $y(x;x') \sim \O(\eta) \O(\eta') H^2
(x-x')^2$ vanishes at coincidence, only the $n=1$ term can contribute after
the action of two derivatives,\footnote{Since the time-ordering commutator
term should not really be present we avoid the delta function term by first 
evaluating for $x^{\mu\prime} \neq x^{\mu}$ and then taking coincidence. 
Another way of getting the same result would be to use the $-+$ propagator.}
\be
\dd_{\r} \dd_{\s}^{\prime} {i\Delta(x;x')}\Bigl\vert_{x' \rightarrow x} = -{H^4 
\over 2^5 \pi^2} \left({H \over \sqrt{\pi}}\right)^{-\epsilon} {\G(4-\epsilon) 
\over 2 - \frac{\e}2} g_{\r\s}(x) \; .
\ee
Contracting with the tensor prefactor in (\ref{kinetic}) gives the one loop
kinetic energy contribution,
\be
{H^4 \over 2^5 \pi^2} \left({H \over \sqrt{\pi}}\right)^{-\epsilon} 
{(1 - \frac{\e}2) \G(4-\epsilon) \over 2 - \frac{\e}2} g_{\mu\nu}(x) \; .
\ee
Although ultraviolet finite, this can be nulled by the following cosmological 
counterterm,
\be
\d\Lambda = {G H^4 \over 4 \pi} \left({H \over \sqrt{\pi}}\right)^{-\epsilon} 
{(1 - \frac{\e}2) \G(4-\epsilon) \over 2 - \frac{\e}2} + O(\lambda) \; .
\ee

\begin{figure}
\centerline{\epsfxsize=0.6\textwidth\epsffile{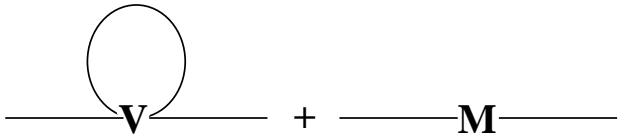}}
\caption{The scalar self mass-squared at order $\lambda$. $V$ denotes the 
4-point vertex and $M$ stands for the mass counterterm vertex.}
\end{figure}

Fig.~1 depicts the one loop scalar self mass-squared:
\be
-i M^2_{\mbox{{\tiny 1-loop}}}(x;x')= -i \left[\frac{\l}2 i\D(x;x) + \d m^2
\right] \d^D(x-x') \; .
\ee
It is calculated by using the coincidence limit of the propagator given in Eq. 
(\ref{coinc}). Because of the finite, time-dependent term we cannot make the
self mass-squared vanish for all time. A reasonable renormalization condition
is that it should be zero at $t=0$. This can be enforced by,
\be
\d m^2=-\frac{\l H^{2-\e}}{2^4\pi^{2-{\e /2}}} \frac1{\e} \G\left(3 - 
\frac{\e}2\right) +O(\l^2) \; . \label{massct}
\ee
The resulting renormalized self mass-squared is,
\be
M^2_{\mbox{{\tiny 1-loop}}}(x;x') = \frac{\l}{8\pi^2} H^2 \, Ht \delta^D(x-x')
\; .
\ee

The temporal growth of $M^2$ has a straightforward physical interpretation. 
It is only at $\phi = 0$ that the scalar potential has zero curvature; for a 
general field configuration the curvature goes like $\phi^2$. But we know from 
(\ref{coinc}) that the expectation value of $\phi^2$ grows with time as the
scalar executes a drunkard's walk. Hence the mass-squared must exhibit the
same time dependence.

\section{The potential energy contributions} \label{section:LD}

\begin{figure}
\centerline{\epsfxsize=0.6\textwidth\epsffile{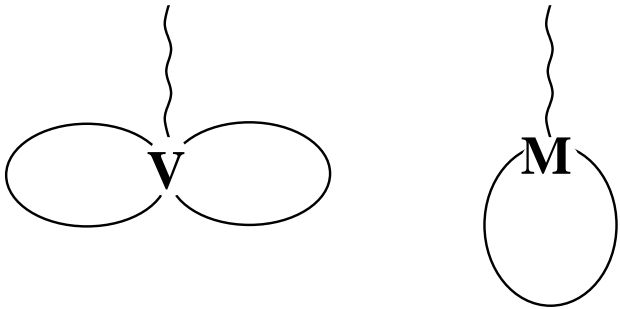}}
\caption{The potential energy contributions to the scalar stress-energy 
tensor at order $\lambda$. $V$ denotes the 4-point vertex and $M$ stands 
for the mass counterterm vertex.}
\end{figure}

The two loop contribution from the potential energy is depicted in Fig.~2.
A simple application of the Feynman rules reveals it to be,
\be
T_{\mu\nu}^{P}(x) = - g_{\mu\nu}(x) \Biggl\{ \frac{\l}8 \Bigl[i \D(x;x)
\Bigr]^2 + \frac{\d m^2}2 i\D(x;x) \Biggr\}\; .
\ee
Substituting our results (\ref{coinc}) for the coincidence limit and 
(\ref{massct}) for the mass counterterm gives,
\be
T_{\mu\nu}^{P}(x) = g_{\mu\nu}(x) \frac{\l H^4}{2^6\pi^4} \Biggl\{\left( 
\frac{\pi}{H^2} \right)^\e \frac{\G^2 (3 - \frac{\e}2)}{8\e^2} - \frac12 
\ln^2\Bigl( \O(\eta)\Bigr) \Biggr\}\; .
\ee
Since $T_{00} \equiv -\r g_{00}$, and $T_{ij} \equiv p g_{ij}$, the
energy density $\r$ and pressure $p$ are,
\be
\r_{\tiny P} = - p_{\tiny P} = \frac{\l H^4}{2^6 \pi^4} \Biggl\{- \left(
\frac{\pi}{H^2} \right)^\e \frac{\G^2(3 - \frac{\e}2)}{8\e^2} + \frac12
\ln^2\Bigl( \O(\eta)\Bigr)\Biggr\}\; . \label{rpL}
\ee
The temporal growth of these terms has the same physical interpretation as
that of the mass-squared. Since the scalar changes slowly we expect that the 
kinetic energy contributions are subdominant. The next section will confirm
this expectation.

\section{Order $\lambda$ kinetic contributions} \label{section:NLD}

Fig.~3 depicts the order $\lambda$ contributions from the kinetic energy. It
is only these graphs which require the full Schwinger-Keldysh formalism, 
very readable derivations of which exist in the literature 
\cite{Schwinger,Jordan}. The rules themselves are simple. For every kind of
vertex that might appear in a Feynman diagram the Schwinger-Keldysh formalism
has two kinds of vertices: a $+$ vertex which is the same as that of 
Feynman, and a $-$ vertex which is the conjugate. Propagation can take place 
between any two kinds of vertices using the appropriate propagator. Each 
propagator is the same function of the four variables $y_{\pm \pm}(x;x')$ 
given at the end of Section 3. 

Using the Schwinger-Keldysh formalism is straightforward. One simply draws the
analogous Feynman diagram and then sums over $\pm$ variations. If the operator
under study is to be time-ordered (as in our case) then the external lines
are $+$. For the diagrams of Fig.~3 this means that the vertex marked $D$ is
fixed to be $+$, however one must sum over both $+$ and $-$ contributions in
the $V$ and $M$ vertices, of course using the appropriate propagators. The
result is,
\begin{eqnarray}
\lefteqn{T_{\mu\nu}^{K}(x) = \left[\delta_\mu\,^\rho \delta_\nu\,^\sigma 
- \frac12 \eta_{\mu\nu} \eta^{\rho\sigma} \right]\! \int d^Dx' \Omega^D(\eta') 
\Big\{\partial_{\r} i\Delta_{++}(x;x') \partial_{\s} i\Delta_{++}(x;x') }
\nonumber \\
& & - \partial_{\r} i\Delta_{+-}(x;x') \partial_{\s} i\Delta_{+-}(x;x')
\Big\} \times \left\{- {i\over 2} \l i\D (x';x') - i \d m^2\right\}\; . 
\label{ana}
\end{eqnarray} 
Since both diagrams of Fig.~3 have the same upper loop, they possess a
common factor in the first curly bracket. The first term within the final 
curly bracket derives from the left hand diagram, while the second term 
comes from the right hand diagram.

\begin{figure}
\centerline{\epsfxsize=0.6\textwidth\epsffile{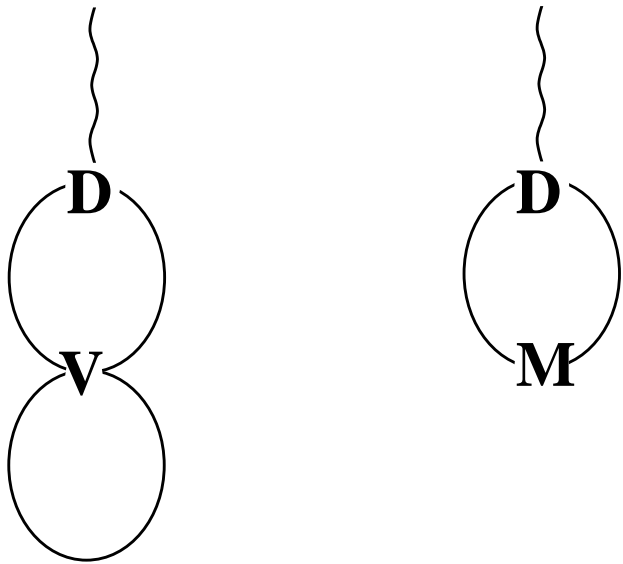}}
\caption{Order $\lambda$ kinetic energy contributions to the scalar 
stress-energy tensor. The derivative vertex is $D$, $V$ denotes the 4-point 
vertex and $M$ stands for the mass counterterm vertex.}
\end{figure}

It is convenient to subsume the complicated, $\e$-dependent constants which
appear in the propagator (\ref{prop}),
\be
i\D(x;x') \equiv \alpha \Bigl\{\g \left( y(x;x')\right) + \beta\ln\Bigl(
\O(\eta) \O(\eta')\Bigr) \Bigr\}\; . \label{propshort}
\ee
Comparison with (\ref{prop}) reveals,
\be
\alpha \equiv \left(\frac{H}{2\pi} \right)^2 \left(\frac{H}{\sqrt{\pi}}
\right)^{-\epsilon} \G (2 - \frac{\epsilon}{2})\; , \hskip 0.7cm \beta
\equiv \frac14 \frac{\G (3-\epsilon)}{\G (2 - \frac{\epsilon}2)} \; ,
\ee
and,
\be
\gamma(y) \equiv \frac1{1 - {\epsilon \over 2}} \frac1{y^{1 - {\epsilon
\over 2}}} + (1-{\epsilon \over 4}) \left(\frac{1 - y^{\frac{\epsilon}{2}}}{
\epsilon}\right) + \sum_{n =1}^{\infty} \Bigg[\frac1{n} {\Gamma(3 + n - 
\epsilon) \over \Gamma(2 + n - \frac{\epsilon}2)} \frac{y^n}{4^{n+1}} 
\nonumber \\
- {1 \over {n + \frac{\epsilon}2}} {{\Gamma(3 + n - \frac{\epsilon}{2})} 
\over {(n+1)! \Gamma(2 - \frac{\epsilon}2)}}} {{y^{n + \frac{\epsilon}2}
\over {4^{n+1}}}\Bigg]\; .
\ee
In this notation, the terms of the last bracket in Eq. (\ref{ana}) become,
\be
-{i\over 2}\l i\D (x';x')-i\d m^2 = -i\l \a\b\ln\left(\O(\eta')\right)\; .
\ee
Derivatives of the propagators can be written as,
\be
\dd_{\r} i\D(x;x') = \alpha \Big\{\gamma'(y)\frac{\dd y}{\dd x^\rho} +
\beta\delta_\rho\,^0 H\O(\eta)\Big\}\; . \label{derprop}
\ee
These can be further reduced by noting,
\be
\frac{\dd y}{\dd x^\rho} = H \O(\eta) \d_\rho\,^0y + 2 H^2 \O(\eta) \O(\eta')
\D x_\rho \; .
\ee
At length one finds,
\be
\dd_{\r} i \D(x;x') \dd_{\s} i\D_(x;x') = \alpha^2 \Bigg\{4
\O^2(\eta) \O^2(\eta') H^4 \D x_\rho\D x_\s {\g '}^2 \nonumber\\
+ 2 \O^2(\eta) \O(\eta') H^3 \Big[\d_\r\,^0\D x_\s +\d_\s\,^0 \D x_\r\Big] 
\Big[y\g'^2 + \b \g'\Big] \nonumber \\
+ \O^2(\eta) H^2 \d_\r\,^0 \d_\s\,^0 \left[ y^2 \g'^2 + 2 \b \g'y + \b^2
\right]\Bigg\}\; . \label{2prop}
\ee

Expression (\ref{2prop}) seems complicated due to the infinite sum in
the definition of $\gamma (y)$. However, we need only retain terms which 
survive as $\epsilon\rightarrow 0$,
\be
\lim_{\epsilon \rightarrow 0} \g'(y) & = & -\frac1{y^2} - \frac{1}{2y}\; ,\\
\lim_{\epsilon \rightarrow 0} \Bigl(\g'(y)\Bigr)^2 & = & \frac1{y^{4 - 
\epsilon}} + \frac{(2-\frac{\epsilon}2)} {2y^{3 - \epsilon}} + \frac1{4y^2}\; .
\ee
One can therefore reduce (\ref{2prop}) to,
\be
\lefteqn{ \a^2 H^2 \O(\eta)^2 \Bigg\{H^2 \O(\eta')^2 \D x_\r\D x_\s\Big[
\frac4{y^{4 - \epsilon}} + \frac{2(2-\frac{\epsilon}2)} {y^{3-\epsilon}}
+ \frac{1}{y^2} \Big]} \nonumber \\
& & + H\O(\eta') \Big[\d_\r\,^0 \D x_\s + \d_\s\,^0 \D x_\r\Big]\Big[
\frac2{y^{3 - \epsilon}} + {1 \over y^2} \Big] + \d_\r\,^0 \d_\s\,^0 
\frac1{y^{2 - \epsilon}}\Bigg\}\; .
\ee
Note that we have retained the regularization for terms which can produce
divergences in $T^K_{\mu\nu}(x)$.

It is useful to break $T^K_{\mu\nu}(x)$ into a sum of six terms of the general
form,
\begin{eqnarray}
\lefteqn{I^n_{\mu\nu}(\eta) \! \equiv \! -i\l \a^3 \b H^2 \O^2(\eta)
\left(\eta_{\mu\rho} \eta_{\nu\sigma} - \frac12 \eta_{\mu\nu} \eta_{\r\s}
\right)} \nonumber \\
& & \times \int d^Dx' \O^{4-\e}(\eta') \ln\left(\O(\eta')\right) 
F_n^{\r\s}(x,x') \; . \label{nonl}
\end{eqnarray}
The functions $F_n^{\r\s}(x,x')$ are defined as follows:
\begin{eqnarray}
F_1^{\r\s}(x,x') & =& 4 H^2 \O(\eta')^2 \D x^\r\D x^\s \Bigg[\frac1{y_{++}^{4
-\e}}-\frac{1}{y_{+-}^{4-\e}}\Bigg]\; ,\\
F_2^{\r\s}(x,x') & = & 2 (2 - \frac{\e}2) H^2 \O(\eta')^2\D x^\r\D
x^\s\Bigg[\frac{1}{y_{++}^{3-\e}}-\frac{1}{y_{+-}^{3-\e}}\Bigg]\; ,\\
F_3^{\r\s}(x,x') & = & H^2\O(\eta')^2\D x^\r\D
x^\s\Bigg[\frac{1}{y_{++}^2}-\frac{1}{y_{+-}^{2}}\Bigg]\; ,\\
F_4^{\r\s}(x,x') & = & 2H\O(\eta')\left[\d^\r\,_0\D x^\s+\d^\s\,_0\D
x^\r\right]\Bigg[\frac{1}{y_{++}^{3-\e}}-\frac{1}{y_{+-}^{3-\e}}\Bigg]\; ,\\
F_5^{\r\s}(x,x') & = & H\O(\eta')\left[\d^\r\,_0\D x^\s+\d^\s\,_0\D
x^\r\right]\Bigg[\frac{1}{y_{++}^2}-\frac{1}{y_{+-}^2}\Bigg]\; ,\\
F_6^{\r\s}(x,x') & = & \d^\r\,_0\d^\s\,_0\Bigg[\frac{1}{y_{++}^{2-\e}}
-\frac{1}{y_{+-}^{2-\e}}\Bigg]\; .
\end{eqnarray}
In the remainder of this section we will evaluate $I^1_{\mu\nu}(\eta)$ 
explicitly to illustrate the relevant techniques, and then simply give the
final answers for the remaining five.

By removing the factors of $\O$ and $H$ the first integral assumes the form,
\be
I^1_{\mu\nu}(\eta) = -\frac{i\l H^{2-\e}}{2^6\pi^{6 - \frac32 \e}} 
\O^{-2+\e}(\eta) \G^2\Bigl(2 - \frac{\e}2\Bigr) \G(3-\e) \left[\d_\mu\,^\r 
\d_\nu\,^\s - \frac12 \eta_{\mu\nu} \eta^{\r\s} \right] \nonumber \\
\times \int d^Dx'\ln\left( \O(\eta') \right) \O^2(\eta') \left\{\frac{\D 
x_\r\D x_\s}{\D x_{++}^{8 - 2\e}} - \frac{\D x_\r\D x_\s}{\D x_{+-}^{8-2\e}}
\right\}\; .  \label{I1}
\ee
Here, $\D x_{++}^{2}\equiv\Vert \vec{x} - \vec{x}' \Vert^2 - (\vert \eta - 
\eta' \vert - i e)^2$, whereas $\D x_{+-}^{2}\equiv\Vert \vec{x} - \vec{x}' 
\Vert^2 - (\eta - \eta' + i e)^2$. Now use the differential identities,
\be
\dd_\r\dd_\s\frac{1}{\D x^{4-2\e}} & = & -(4-2\e)\left[ \frac{\eta_{\r\s}}{\D
x^{6-2\e}} - (6-2\e) \frac{\D x_\r\D x_\s}{\D x^{8-2\e}}\right]\; , \\
\dd^2 \frac1{\D x^{4-2\e}} & = & 2(2-\e)^2 \frac1{\D x^{6-2\e}}\: , \\
\dd^2 \frac1{\D x^{2-2\e}} & = & -2\e (1-\e) \frac1{\D x^{4-2\e}}\: ,
\ee
to write,
\be
\frac{\D x_\r\D x_\s}{\D x^{8-2\e}} = -\frac1{8\e(1-\e)(2-\e)(3-\e)}
\left[\dd_\r\dd_\s + \frac1{2-\e} \eta_{\r\s} \dd^2\right] \dd^2
\frac1{\D x^{2-2\e}} \; .
\ee
These results pertain for both the $++$ and $+-$ terms. Since the range of
$x^{\mu\prime}$ integration does not depend upon $x^{\mu}$, the derivatives
can be pulled outside,
\begin{eqnarray}
\lefteqn{I^1_{\mu\nu}(\eta) = \frac{i\l H^{2-\e}}{2^9 \pi^{6 - \frac32 \e}} 
\O^{-2+\e}(\eta) {\G^2\Bigl(2 - \frac{\e}2\Bigr) \G(2-\e) \over (1-\e)
(3-\e)} \left[\dd_{\mu} \dd_{\nu} - \eta_{\mu\nu} \dd^2 \right] }
\nonumber \\
& & \hspace{2cm} \times \int d^Dx' \ln\left( \O(\eta')\right) \O^2(\eta')
{\dd^2 \over \e} \left[ \frac{1}{\D x_{++}^{2-2\e}} - \frac1{\D x_{+-}^{2-
2\e}} \right] \; . \label{Int}
\end{eqnarray}

It is at this stage that the parallel treatment of the $++$ and $+-$ ends.
The crucial difference between the two terms is that $\D x_{++}$ contains an
absolute value, $|\eta-\eta'|$, whereas $\D x_{+-}$ does not. Double 
derivatives of $x_{++}$ can therefore result in a temporal delta function.
Taking $e$ to zero may or may not produce a factor of $\d^D(x - x')$, 
depending upon how many powers of $x_{++}$ there are. The unique power for 
general dimension $D = 4 -\e$ turns out to be,
\be
\dd^2 \frac1{\D x_{++}^{2-\e}} = {2 e (2-\e) i\d (\D\eta) \over (\Vert 
\vec{x} - \vec{x}' \Vert + e^2)^{2 -\frac{\e}{2}}} \rightarrow  4i 
\frac{\pi^{2 - \frac{\e}2}}{\G(1 - \frac{\e}2)} \d^D(x-x') \label{2-e++} \; ,
\ee
In contrast the $+-$ term gives zero,
\be
\dd^2 \frac1{\D x_{+-}^{2-\e}} = 0 \; . \label{2-e+-}
\ee

By using (\ref{2-e+-}) we can write the $+-$ term in (\ref{Int}) in a form
that remains finite in the unregulated limit,
\be
{\dd^2 \over \e} \left[\frac1{\D x^{2-2\e}_{+-}}\right] = {\dd^2 \over \e}
\left[\frac{1}{\D x^{2-2\e}_{+-}} - \frac{\mu^{-\e}}{\D x^{2-\e}_{+-}} 
\right] \; . \label{sub}
\ee
The arbitrary mass parameter $\mu$ has been introduced to maintain the correct 
dimensions. Since this term is manifestly ultraviolet finite we may as well
take $\e$ to zero,
\be
{\dd^2 \over \e} \left[\frac{1}{\D x^{2-2\e}_{+-}}\right] \rightarrow \dd^2 
\left[ \frac{\ln\left(\mu^2\D x^2_{+-}\right)}{2 \D x^2_{+-}} \right]\; . 
\label{Dx2+-}
\ee
The analogous reduction of the $++$ term gives,
\begin{eqnarray}
{\dd^2 \over \e} \frac{1}{\D x_{++}^{2-2\e}} & = & {\dd^2 \over \e} \left[
\frac1{\D x_{++}^{2-2\e}} - \frac{\mu^{-\e}}{\D x_{++}^{2-\e}}\right] + 
\frac{4i \pi^{2- \frac{\e}2} \mu^{-\e}}{\e \G(1 - \frac{\e}2)} \d^D(x-x') 
\; , \\
& \rightarrow & \dd^2\left[\frac{\ln\left(\mu^2\D x^2_{++}\right)}{2 \D 
x^2_{++}} \right] + \frac{4i \pi^{2 - \frac{\e}2} \mu^{-\e}}{\e \G(1 - 
\frac{\e}2)} \d^D(x-x')\; . \label{Dx2++}
\end{eqnarray}

Using (\ref{Int}), (\ref{Dx2+-}) and (\ref{Dx2++}) we can bring the first
integral to the form,
\be
\lefteqn{I^1_{\mu\nu}(\eta) = \frac{i\l H^{2-\e} \O^{-2 + \e}(\eta)}{2^9\pi^{
6 - \frac32 \e}} \frac{\G(2-\e) \G^2(2 - \frac{\e}2)}{(1 - \e) (3-\e)} \left[ 
\dd_\mu\, \dd_\nu - \eta_{\mu\nu} \dd^2\right]} \nonumber \\
& & \times \int d^dx'\ln\left( \O(\eta')\right) \O(\eta')^2 \Bigg\{\dd^2
\left[\left(\frac{\ln\left( \mu^2 \D x^2_{++} \right)}{2 \D x^2_{++}}
- \frac{\ln\left(\mu^2\D x^2_{+-} \right)}{2 \D x^2_{+-}} \right)\right]
\nonumber \\
& & + {4i \pi^{2 - \frac{\e}2} \mu^{-\e} \over \e \G(1 - \frac{\e}2)} 
\d^D(x-x') \Bigg\}\; . 
\ee
Now note that the integral has no dependence upon $\vec{x}$, so only the
temporal derivatives matter. Evaluating $\d$-function integral and taking 
$\epsilon$ to zero in the finite terms gives,
\be
\lefteqn{I^1_{\mu\nu}(\eta) = -\frac{\l H^{4-\e} \O^{2+\e}(\eta)}{2^7 \pi^{4-
\e}}\, \frac{(1 - \frac{\e}2) \G(1 - \e) \G(2 - \frac{\e}2) \mu^{-\e}}{\e
(3-\e)}\left[ \d_\mu\,^0 \, \d_\nu\,^0 + \eta_{\mu\nu} \right]} \nonumber \\
& & \times \left[6\ln \left( \O(\eta)\right) + 5)\right]
+\frac{i\l H^2 \O^{-2}(\eta)}{2^{10} 3\pi^6} \left[\d_\mu\,^0 \,\d_\nu\,^0
+ \eta_{\mu\nu} \right] \nonumber \\
& & \times \dd_0^4 \int d^Dx'\ln\left( \O(\eta')\right) \O^2(\eta') \Bigg\{
\frac{\ln\left( \mu^2 \D x^2_{++}\right)}{\D x^2_{++}} - \frac{\ln\left(
\mu^2 \D x^2_{+-} \right)}{\D x^2_{+-}} \Bigg\}\; . \label{I11}
\ee

To evaluate the integral in Eq. (\ref{I11}) we pull out another derivative
using the identities:
\be
\dd^2 \ln^2\left( \D x^2\right) & = & 8\left[\frac{ \ln\left(\D x^2 \right)}{
\D x^2} + \frac1{\D x^2}\right] \;, \\
\dd^2 \ln\left(\D x^2\right) & = & \frac4{\D x^2} \; .
\ee
The remaining integrand possesses only logarithmic singularities. If we define
the coordinate separations,
\be
\D\eta \equiv \eta - \eta' \qquad , \qquad r \equiv \Vert \vec{x} - \vec{x}'
\Vert \; ,
\ee
the logarithms can be expanded as,
\begin{eqnarray}
\ln\left[ \mu^2\D x^2_{++} \right] & = & \ln\left[ \mu^2 (\D\eta^2 - r^2)
\right] + i\pi \theta(\D \eta^2 - r^2) \; , \\
\ln\left[ \mu^2 \D x^2_{+-} \right] & = & \ln\left[ \mu^2 (\D \eta^2 - r^2)
\right] - i\pi \theta( \D \eta^2 - r^2) \; .
\ee
Putting it all together gives the following reduction,
\be
\lefteqn{\dd_0^4 \int d^Dx'\ln\left( \O(\eta') \right) \O^2(\eta') \Bigg\{
\frac{\ln\left( \mu^2 \D x^2_{++} \right)}{\D x^2_{++}} - \frac{\ln\left(
\mu^2 \D x^2_{+-} \right)}{\D x^2_{+-}} \Bigg\}} \nonumber \\
& & = -i2 \pi^2 \dd_0^6 \int_{-{1 \over H}}^\eta d\eta' \ln\left( \O(\eta')
\right) \O^2(\eta') \int_0^{\D\eta} dr r^2 \left(\ln\left[ \mu^2 (\D \eta^2
- r^2)\right] - 1 \right) \nonumber \\
& & = -i2 \pi^2 \dd_0^6 \int_{-{1 \over H}}^\eta d\eta' \ln\left( \O(\eta')
\right) \O^2(\eta') (\D \eta)^3 \left[ {2 \over 3} \ln\left(2 \mu \D\eta
\right) - {11 \over 9} \right] \nonumber \\
& & = -i8 \pi^2 \dd_0^3 \int_{-{1 \over H}}^\eta d\eta' \ln\left( \O(\eta')
\right) \O^2(\eta') \ln\left(2 \mu \D\eta \right)\; .
\ee

The lower limit of temporal integration at $\eta' = -H^{-1}$ (that is, $t' = 
0$) derives from the fact that we release the state in free Bunch-Davies 
vacuum at this instant. $I_1$ can now be recast as:
\be
I^1_{\mu\nu}(\eta) & = & -\frac{\l H^4}{2^6 \pi^4} \O^2(\eta) \left[\d_\mu\,^0
\d_\nu\,^0 + \eta_{\mu\nu} \right] \Bigg\{ \frac{\zeta}{\e} \frac{\O^{\e}(\eta)
}{(3 - \e)} \left[3 \ln\left( \O(\eta)\right) + {5 \over 2}\right] \nonumber \\
& & + \frac1{6H^2 \O^4(\eta)} \dd_0^3 \int_{- {1 \over H}}^\eta d\eta'
\ln\left( \O(\eta') \right) \O(\eta')^2 \ln\left(2 \mu\D \eta\right)\Bigg\}\; ,
\label{I111}
\ee
where $\zeta$ is defined as
\be
\zeta \equiv \left(\frac{\pi}{\mu H} \right)^\e \left(1 - \frac{\e}2\right)^2
\G(1 - \e) \G\left(1 - \frac{\e}2\right)\; .
\ee
To evaluate the integral in Eq. (\ref{I111}), we change variables to
$\O(\eta') = -\frac1{H\eta'}$ and expand the logarithm,
\be
\!\!\!\!\!\!\!\!\!\!\!\! & & \!\!\!\!\!\! \frac1{6H^2\O^4(\eta)} \dd_0^3
\int_{-{1 \over H}}^\eta d\eta'\ln\left( \O(\eta')\right) \O^2(\eta')
\ln\left(2 \mu\D\eta\right) \nonumber \\
\!\!\!\!\!\!\!\!\! & & \!\!\! = \frac1{6 \O^4} \left(\O^2 \frac{\dd}{\dd \O} 
\right)^2 \Biggl[ \O^2 \ln\left(\O \right) \left[\ln( \frac{2\mu}{H}) - 
\ln\left( \O\right) \right]\Biggr] \nonumber \\
\!\!\!\!\!\!\!\!\! & & \!\!\! - \frac1{6\O^4} \left(\O^2 \frac{\dd}{\dd\O}
\right)^3 \sum_n^\infty \frac1{n \O^n} \int_1^{\O} d\O' \O^{\prime n}
\ln\left( \O'\right) \label{dkup} \\
& = & -\ln^2\left( \O\right) - \frac83 \ln\left( \O\right) + \ln\left( 
\frac{2\mu}{H} \right) \ln\left( \O\right) + \frac56 \ln\left(\frac{2\mu}{H} 
\right) \nonumber \\
& & -\frac16 \left[1 + \pi^2 - \sum_n^\infty \frac{(n-1)(n-2)}{(n+1)^2}
\O^{-n-1} \right]\; . \label{dI}
\ee

Our final result for $I^1_{\mu\nu}(\eta)$ is,
\begin{eqnarray}
\lefteqn{I^1_{\mu\nu}(\eta) = - \frac{\l H^4}{2^6 \pi^4} \O^2 \left[\d_\mu\,^0 
\d_\nu\,^0 + \eta_{\mu\nu} \right]\Biggl\{ \Bigg[ \frac{\zeta}{\e} + 
\ln\left(\frac{2\mu}{H} \right) - \frac32 \Bigg] \ln\left(\O \right) }
\nonumber \\
& & + \frac56 \Bigg[\frac{\zeta}{\e} + \ln\left(\frac{2\mu}{H} \right)\Bigg] 
+ \frac19 - \frac{\pi^2}6 + \frac16 \sum_{n=1}^\infty \frac{(n-1) (n-2)}{
(n+1)^2} \O^{-n-1} \Biggr\}\; .
\end{eqnarray}
The five other terms in Eq. (\ref{nonl}) can be evaluated similarly. The
results are,
\begin{eqnarray}
\lefteqn{I^2_{\mu\nu}(\eta) = - \frac{i\l H^{4-\e}}{2^7 \pi^{6 - \frac32\e}} 
\O^{-1 + \e} \G(2 - \frac{\e}2) \G(3-\e) \G(3-\frac{\e}2) \left[\d_\mu \,^\r 
\d_\nu\,^\s - \frac12 \eta_{\mu\nu} \eta^{\r\s} \right] } \nonumber \\
& & \hspace{3cm} \times \int d^Dx'\ln\left( \O(\eta') \right) \O^3(\eta') 
\left\{ \frac{\D x_\r \D x_\s}{\D x_{++}^{6-2\e}} - \frac{\D x_\r \D x_\s}{
\D x_{+-}^{6 - 2\e}} \right\} \nonumber \\
& & \hspace{2cm} = \frac{\l H^4}{2^6 \pi^4} \O^2 \Biggl\{ \eta_{\mu\nu} \left[ 
\Bigg[\frac{\zeta}{\e} + \ln\left( \frac{2\mu}{H} \right) - \frac94 \right] 
\ln\left( \O \right) + \frac54 - \frac{\pi^2}6 \nonumber \\
& & \hspace{3cm} + \sum_{n=2}^\infty \frac{{\O^{-n-1}}}{(n+1)^2} \Biggr] - 
\left[\d_\mu\,^0 \d_\nu\,^0+ \frac12 \eta_{\mu\nu} \right] \ln(\O) \Bigg\}
\; . \\
\lefteqn{I^3_{\mu\nu}(\eta) = - \frac{i\l H^{6-3\e}}{2^8 \pi^{6 - 
\frac32 \e}} \G^2(2-\frac{\e}2) \G(3-\e) \left[\d_\mu\,^\r \d_\nu\,^\s - 
\frac12 \eta_{\mu\nu} \eta^{\r\s} \right] } \nonumber \\
& & \hspace{3cm} \times \int d^Dx' \ln\left( \O(\eta') \right) \O^{4-\e}(\eta') 
\left\{ \frac{\D x_\r \D x_\s}{\D x_{++}^4} - \frac{\D x_\r\D x_\s}{\D 
x_{+-}^4} \right\} \nonumber \\
& & \hspace{2cm} = - \frac{\l
H^4}{2^6\pi^4} \O^2 \d_\mu\,^0 \d_\nu\,^0
\Bigg\{\frac13 \ln\left( \O\right) - \frac5{18} + \frac12 \O^{-2}
- \frac29 \O^{-3} \Bigg\}\; . \\
\lefteqn{I^4_{\mu\nu}(\eta) = - \frac{i\l H^{3-\e}}{2^7\pi^{6 - \frac32 \e}} 
\O^{-1 + \e} \G^2(2 - \frac{\e}2) \G(3-\e) \left[\d_\mu\,^\r \d_\nu\,^\s -
\frac12 \eta_{\mu\nu} \eta^{\r\s} \right] } \nonumber \\
& & \times \int d^Dx' \ln\left( \O(\eta') \right) \O^2(\eta') \left[\d_\r\,^0 
\D x_\s + \d_\s\,^0 \D x_\r \right] \left\{\frac1{\D x_{++}^{6-2\e}} - 
\frac1{\D x_{+-}^{6-2\e}} \right\} \nonumber \\
& & = \frac{\l H^4}{2^6 \pi^4} \O^2(\eta) \left[\d_\mu\,^0 \d_\nu\,^0 + 
\frac12 \eta_{\mu\nu} \right]\Bigg\{ \Bigg[\frac{\zeta}{\e} + \ln\left( 
\frac{2\mu}{H} \right) - \frac32 \Bigg] 2 \ln\left( \O \right) \nonumber \\
& & \hspace{2cm} + \frac{\zeta}{\e} + \ln\left(\frac{2\mu}{H} \right) + 1 -
\frac{\pi^2}3 - \sum_{n=1}^\infty \frac{(n-1)}{(n+1)^2} \O^{-n-1} \Bigg\}\; .\\
\lefteqn{I^5_{\mu\nu}(\eta) = - \frac{i\l H^{5-3\e}}{2^8 \pi^{6 - \frac32 
\e}} \G^2(2 - \frac{\e}2) \G(3-\e) \left[\d_\mu\,^\r \d_\nu\,^\s - \frac12 
\eta_{\mu\nu} \eta^{\r\s} \right] } \nonumber \\
& & \times \int d^Dx'\ln\left( \O(\eta') \right) \O^{3 -\e}(\eta')
\left[\d_\r\,^0 \D x_\s + \d_\s\,^0 \D x_\r\right] \left\{\frac1{\D x_{++}^4} 
- \frac1{\D x_{+-}^4}\right\} \nonumber \\
& & \hspace{2cm} = \frac{\l H^4}{2^6 \pi^4} \O^2 \left[\d_\mu\,^0\d_\nu\,^0 + 
\frac12 \eta_{\mu\nu} \right] \Bigg\{ \ln\left( \O\right) - \frac12 + \frac12 
\O^{-2} \Bigg\} \; . \\
\lefteqn{I^6_{\mu\nu}(\eta) = - \frac{i \l H^{4-\e}}{2^8 \pi^{6 - \frac32 
\e}} \O^\e \G^2(2 - \frac{\e}2) \G(3 - \e) \left[\d_\mu\,^0 \d_\nu\,^0 + 
\frac12 \eta_{\mu\nu} \right] } \nonumber \\
& & \hspace{3cm} \times \int d^Dx' \ln\left( \O(\eta') \right) \O^2(\eta') 
\left\{ \frac1{\D x_{++}^{4 - 2\e}} - \frac1{\D x_{+-}^{4-2\e}} \right\}
\nonumber \\
& & \hspace{2cm} = - \frac{\l H^4}{2^6 \pi^4} \O^2 \left[\d_\mu\,^0
\d_\nu\,^0 + \frac12 \eta_{\mu\nu} \right] \Bigg\{\Bigg[ \frac{\zeta}{\e} + 
\ln\left(\frac{2\mu}{H} \right) - \frac32 \Bigg] \ln\left( \O\right)
\nonumber \\
& & \hspace{5cm} + 1 - \frac{\pi^2}6 + \sum_{n=1}^\infty \frac{\O^{-n-1}}{
(n+1)^2} \Bigg\}\; .
\end{eqnarray}

Summing the six terms gives the total order $\lambda$ contribution from the
kinetic energy,
\be
& & \!\!\!\!\!\!\!\!\!\!\!\!\!\!\!\!\!\!\!\! T_{\mu\nu}^K(x)
= \frac{\l H^4}{2^6 \pi^4} \O^2 \Bigg\{ \d_\mu\,^0 \d_\nu\,^0 \Bigg[\frac16 
\Bigg(\frac{\zeta}{\e} + \ln\left( \frac{2\mu}{H} \right)\Bigg) - \frac13 
\ln\left( \O\right) \nonumber \\
& & \!\!\!\!\!\!\!\!\!\!\!\!\!\! - \frac13 + \frac29 \O^{-3} - \frac16
\sum_{n=1}^{\infty} \frac{(n+2)}{(n+1)} \O^{-n-1} \Bigg] + \eta_{\mu\nu}
\Bigg[\Bigg( \frac{\zeta}{\e} + \ln\left(\frac{2\mu}{H} \right)\Bigg) \times
\nonumber \\
& & \!\!\!\!\!\!\!\!\!\!\!\!\!\!\!\!\!\!\!\! \Bigg(\frac12 \ln\left( \O
\right) - \frac13 \Bigg) - \frac32 \ln\left( \O\right) + \frac89 -
\frac{\pi^2}{12} - \frac16 \sum_{n=1}^{\infty} \frac{(n^2-4)}{(n+1)^2} 
\O^{-n-1} \Biggr] \Bigg\}\; .
\ee
We therefore obtain the following order $\l$ kinetic energy density,
\be
\r_{ \footnotesize K} = \frac{\l H^4}{2^6\pi^4} \Biggl\{\left[ \frac{\zeta}{
\e} + \ln\left( \frac{2\mu}{H} \right)\right] \left[-\frac12 \ln\left(
\O \right) + \frac12 \right] + \frac76 \ln\left( \O\right) - \frac{11}9 
\nonumber \\
+ \frac{\pi^2}{12} + \frac29 \O^{-3} - \frac12 \sum_{n=1}^\infty \frac{(n+2)}{
(n+1)^2} \O^{-n-1} \Biggr\}\; , \label{rhoNL}
\ee
and pressure,
\be
p_{ \footnotesize K} = \frac{\l H^4}{2^6\pi^4} \Biggl\{\left[ \frac{\zeta}{\e}
+ \ln\left(\frac{2\mu}{H} \right)\right] \left[\frac12 \ln\left( \O
\right) - \frac13 \right] - \frac32 \ln\left( \O\right) + \frac89 - 
\frac{\pi^2}{12} \nonumber \\
- \frac16 \sum_{n=1}^\infty \frac{(n^2-4)}{(n+1)^2} \O^{-n-1} \Biggr\}
\; . \label{pNL}
\ee

\section{The conformal counterterm} \label{section:CCT}

Since $\zeta = 1 + O(\e)$, $T^K_{\mu\nu}$ contains a divergence proportional 
to $g_{\mu\nu} \ln(\O)/\e$. This is an overlapping divergence, and it must
be canceled if the stress-energy tensor is to be a well defined operator. 
It cannot be absorbed into a renormalization of the cosmological constant on 
account of the factor of $\ln(\O)$. The difficulty is resolved by the 
conformal counterterm (\ref{conf}),\footnote{The potential problem and its 
resolution were adumbrated in previous work \cite{AbWo}. }
\begin{eqnarray}
\lefteqn{T^C_{\mu\nu}(x) = - 2\delta\xi \left\{ (D-1) H^2 g_{\mu\nu}(x) 
i\Delta(x;x) \right. } \nonumber \\
& & \hspace{.5cm} \left. + \left[g_{\mu\nu}(x) g^{\r\s}(x) - \d^{\r}_{~\mu} 
\d^{\s}_{~\nu} \right] \left( \dd_{\r} \dd_{\s} - \Gamma^{\alpha}_{~\r\s}(x)
\dd_{\alpha} \right) i\Delta(x;x) \right\} . \qquad 
\end{eqnarray}
(See the first of the graphs on Fig.~4.) Recall that this comes from a term 
in the action which is contrived to vanish in de Sitter background so it 
affects only the stress tensor and makes no contribution to purely scalar 
processes.

The affine connection for de Sitter conformal coordinates is,
\be
\G^\r\,_{\mu\nu} = H \O(\eta) \left[\d^\r\,_\mu \d^0\,_\nu + \d^\r\,_\nu
\d^0\,_\mu - \eta_{\mu\nu} \eta^{\r 0} \right]\; .
\ee
Our simple, $D$-dimensional result for the coincidence limit (\ref{coinc}) of
the propagator implies the following result for the covariant derivative,
\be
\left( \dd_{\r} \dd_{\s} \! - \! \Gamma^{\alpha}_{~\r\s}(x) \dd_{\alpha} 
\right) i\Delta(x;x) \! = \! -\frac{H^{4-\e}}{2^3 \pi^{2-\e/2}} \G(3 \! - \!
\e) \left[ g_{\mu\nu} \! + \! \d_\mu\,^0 \d_\nu\,^0 \O^2(\eta) \right] . \quad
\ee
Substitution and a few trivial manipulations yield,
\begin{eqnarray}
\lefteqn{T_{\mu\nu}^C(x) = - \frac{\d\xi \, H^{4-\e}}{2^2\pi^{2-\e/2}} 
\Bigg\{\Bigg[ \left(\frac{3-\e}{\e} \right) \G\left(3-\frac{\e}2\right) + 
\G(4-\e) \left[\ln\Bigl(\O \right) - 1\Bigr] \Bigg] g_{\mu\nu} } \nonumber \\
& & \hspace{6cm} + \G(3-\e)\left[ g_{\mu\nu} + \d_\mu\,^0 \d_\nu\,^0 \O^2
\right]\Bigg\} . \qquad
\end{eqnarray}
It is straightforward to extract the energy density and pressure,
\begin{eqnarray}
\r_C\!\!\! & = & \!\!\! \frac{\d\xi\,H^{4-\e}}{2^2\,\pi^{2-\e/2}} \Bigg\{
\left(\frac{3-\e}{\e} \right) \G\left(3 - \frac{\e}2\right) + \G(4-\e) 
\Bigl(\ln \left( \O\right) - 1\Bigr)\Bigg\}\; , \\
p_C\!\!\! & = &\!\!\!- \frac{\d\xi\,H^{4-\e}}{2^2\,\pi^{2-\e/2}} \Bigg\{
\left(\frac{3-\e}{\e} \right) \G\left(3-\frac{\e}2\right) \nonumber \\
& & \hspace{4cm} + \G(4-\e) \Bigl( \ln\left( \O)\right) - 1\Bigr) + 
\G(3-\e)\Bigg\}\; .
\end{eqnarray}

\begin{figure}
\centerline{\epsfxsize=0.6\textwidth\epsffile{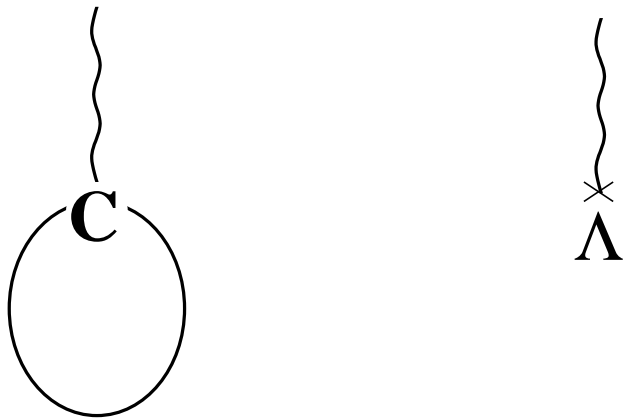}}
\caption{Contributions to the scalar stress-energy tensor at order $\lambda$
from the conformal counterterm $C$ and the counterterm for the bare
cosmological constant $\Lambda$.}
\end{figure}

\section{The fully renormalized result} \label{section:FR}

Summing the potential and kinetic energy densities (\ref{rpL},\ref{rhoNL})
gives,
\be
\rho_{PK} & = & \frac{\l H^4}{2^6\pi^4} \Bigg\{-\left( \frac{\pi}{H^2}\right)^{
\e} \frac{\G^2 (3 - \frac{\e}2)}{8\e^2} + \left[\frac{\zeta}{\e} + \ln\left(
\frac{2\mu}{H} \right)\right] \left[\frac12 - \frac12 \ln\left(\O(\eta)
\right)\right] \nonumber\\
& & + \frac12 \ln^2\left( \O(\eta) \right) + \frac76 \ln\left( \O(\eta)\right)
- \frac{11}9 + \frac{\pi^2}{12} + \frac29 \O(\eta)^{-3} \nonumber \\
& & - \frac12 \sum_{n=1}^\infty \frac{(n+2)}{(n+1)^2} \O(\eta)^{-n-1}
\Bigg\}\; , \label{rhoT}
\ee
Doing the same for the potential and kinetic pressures (\ref{rpL},\ref{pNL})
yields,
\be
p_{PK} & = & \frac{\l H^4}{2^6\pi^4} \Bigg\{\left( \frac{\pi}{H^2} \right)^{\e}
\frac{\G^2 (3 - \frac{\e}2)}{8\e^2} + \left[\frac{\zeta}{\e} + \ln\left(
\frac{2\mu}{H} \right)\right]\left[ \frac12 \ln\left(\O(\eta) \right) - 
\frac13 \right] \nonumber \\ 
& & - \frac12 \ln^2\left( \O(\eta)\right) - \frac32 \ln\left(\O(\eta) \right)
+ \frac89 - \frac{\pi^2}{12} \nonumber \\
& & - \frac16 \sum_{n=1}^\infty \frac{(n^2-4)}{(n+1)^2} \O(\eta)^{-n-1}
\Bigg\} \; . \label{pT}
\ee
It is an important check of the calculation that (\ref{rhoT}) and (\ref{pT}) 
do satisfy the D-dimensional conservation law:
\be
\dot{\rho}_{PK} = -(3-\e) H (\rho_{PK} + p_{PK})\; .
\ee

Although conserved, the potential and kinetic terms still harbor divergences.
We eliminate the nonlocal, ``overlapping'' divergence by choosing the divergent
part of the conformal counterterm thus,
\be
\d\xi \equiv \frac{\l}{2^4\pi^2} \left(\frac{H^2}{\pi} \right)^{\e/2}
\Bigg\{ \frac{\zeta}{2\e\G(4 - \e)} + \d\xi_{\mbox{\tiny{fnt}}}\Bigg\} \; .
\ee
We will choose the finite part $\d\xi_{\mbox{\tiny{fnt}}}$ later so as to
achieve maximum simplicity in the fully renormalized result. The remaining 
divergences are proportional to $g_{\mu\nu}$ so they can be absorbed by a 
cosmological counterterm. The natural choice is,
\be
\frac{\d\Lambda}{8\pi G} = \frac{\l H^4}{2^6\pi^4} \Bigg\{\left( \frac{\pi}{
H^2} \right)^{\e} \frac{\G^2(3-\frac{\e}{2})}{8 \e^2} - \frac{\zeta}{2\e^2}
\frac{\G(3-\frac{\e}{2})}{\G(3-\e)} \nonumber \\
-\left(\frac{3 - \e}{\e} \right) \G\left(3 - \frac{\e}2\right) \d\xi_{
\mbox{\tiny{fnt}}} + \d\Lambda_{\mbox{\tiny{fnt}}}\Bigg\}\; ,
\ee
where we leave the finite part for later determination.

Because the conformal and cosmological counterterms cancel all ultraviolet
divergences we can take $\e$ to zero in the renormalized energy density and
pressure,
\begin{eqnarray}
\lefteqn{\rho_{\mbox{\tiny{ren}}} = \frac{\l H^4}{2^6\pi^4} \Bigg\{\frac12
\ln^2\left(\O\right) + \left[\frac76 - \frac12 \ln\left(\frac{2\mu}{H}
\right) + 6\d\xi_{\mbox{\tiny{fnt}}}\right] \ln\left(\O\right) -\frac{11}9 
+ \frac{\pi^2}{12}} \nonumber\\
& & + \frac12 \ln\left(\frac{2\mu}{H}\right) - 6\d \xi_{\mbox{\tiny{fnt}}}
+ \d\Lambda_{\mbox{\tiny{fnt}}} +\frac29 \O(\eta)^{-3} - \frac12 
\sum_n^{\infty} \frac{n+2}{(n+1)^2} \O^{-n-1}\Bigg\} \; , \qquad \\
\lefteqn{p_{\mbox{\tiny{ren}}} = -\frac{\l H^4}{2^6\pi^4} \Bigg\{\frac12 
\ln^2\left( \O\right) + \left[\frac32 - \frac12 \ln\left(\frac{2\mu}{H} \right)
+ 6\d\xi_{\mbox{\tiny{fnt}}} \right] \ln\left( \O\right) -\frac56 } \nonumber\\
& & + \frac{\pi^2}{12} + \frac13 \ln\left(\frac{2 \mu}{H}\right) -4 \d\xi_{
\mbox{\tiny{fnt}}} + \d\Lambda_{\mbox{\tiny{fnt}}} + \frac16 \sum_n^{\infty}
\frac{(n^2-4)}{(n+1)^2} \O^{-n-1}\Bigg\}\; . \qquad
\end{eqnarray}
The result can be considerably simplified by making a suitable choice for the
finite parts of the counterterms,
\be
\d\xi_{\mbox{\tiny{fnt}}} = -\frac{7}{36} + \frac1{12} \ln\left(\frac{2\mu}{H}
\right) \qquad , \qquad \d\Lambda_{\mbox{\tiny{fnt}}} = \frac1{18} -
\frac{\pi^2}{12} \; .
\ee
The final results are,
\begin{eqnarray}
\r_{\mbox{\tiny{ren}}} & = & \frac{\l H^4}{2^6\pi^4}\Bigg\{ \frac12 
\ln^2\left( \O\right) + \frac29 \O^{-3} - \frac12 \sum_{n=1}^{\infty} 
\frac{n+2}{(n+1)^2} \O^{-n-1} \Bigg\} \; ,\\
p_{\mbox{\tiny{ren}}} & = & - \frac{\l H^4}{2^6\pi^4} \Bigg\{\frac12 
\ln^2\left(\O \right) + \frac13 \ln\left( \O\right) + \frac16 
\sum_{n=1}^{\infty} \frac{n^2-4}{(n+1)^2} \O^{-n-1}\Bigg\}\; .
\end{eqnarray}

We have verified the conjecture of ref. \cite{AbWo} that the kinetic
contributions are subdominant to the potential contributions by one factor
of $\ln(\O)$. This suffices to prove that the model violates the Weak Energy
Condition on cosmological scales,
\be
\rho_{\mbox{\tiny{ren}}} + p_{\mbox{\tiny{ren}}} = \frac{\l H^4}{2^6\pi^4} 
\Bigg\{- \frac13 \ln\left( \O\right) + \frac29 \O^{-3} - \frac16 
\sum_{n=1}^{\infty} \frac{n+2}{n+1} \O^{-n-1}\Bigg\}\; . \label{it}
\ee
The leading term of this expression is exactly that predicted in equation (49)
of ref. \cite{AbWo}. The difference is that we have rigorously proved it. We
also have the sub-dominant corrections {\it and} we have a procedure that can
be pushed to arbitrarily high order.

Two comments are in order concerning (\ref{it}). First, there is a logarithmic
singularity at $\ln(\O) = Ht = 0$ whereas the correct result is that the
stress-energy tensor vanishes then. The problem seems to be caused by assuming
$\ln(\O) \gg \epsilon$ at various places in the reduction. This means we
should not trust the result for times infinitesimally close to $H t=0$ but it
should be completely reliable at later times.

The second comment has to do with the length of time during which we can expect
the Weak Energy Condition to be violated. The $\phi^4$ potential obviously has
some tendency to force the field back down. One might expect this classical
effect to eventually balance against the quantum uncertainty pressure to result
in a constant energy density and pressure.\footnote{However, one should be
alive to the possibility that stochastic effects result in the scalar's further
migration up its potential in certain portions of the universe \cite{Linde}.}
We can estimate how long the system takes to reach this stage by asking
when higher order effects become comparable with the two loop result.

Consider a diagram with $2N$ external scalar lines. At $L$ loop order the 
number of $\phi^4$ interaction vertices is,
\begin{equation}
V = L + N - 1 \; .
\end{equation}
Each contributes a factor of $\lambda$, so the diagram goes like $\lambda^{L
+ N - 1}$. The number of internal propagators is,
\begin{equation}
P = 2L + N - 2 \; .
\end{equation}
Since we are computing a Schwinger diagram, there will be $V$ cancellations 
between $+$ and $-$ variations, which give the $\theta$-function imaginary 
part of the logarithm. However, there are also $V$ temporal integrations, 
each one of which can potentially result in an extra factor of $\ln (\O)$.
Hence the strongest possible effect for the $2N$-point vertex at $L$ loop 
order is,
\begin{equation}
V_{2N}^L \sim \lambda^{L + N - 1} \Bigl(\ln(\O)\Bigr)^{2L + N - 2} \; .
\end{equation}
The stress-energy tensor corresponds to $N=0$ scalar lines so the dominant 
contribution at $L$ loop order is,
\begin{equation}
T_{\mu\nu}^{\tiny L-loop} \sim g_{\mu\nu} H^4 \left(\lambda \ln^2(\O)
\right)^{L-1} \; .
\end{equation}
It follows that perturbation theory breaks down when $\ln(\O) \sim 
1/\sqrt{\lambda}$. Since $\lambda$ is assumed small we see that the phase of
super-acceleration can be made to last for an enormous number of e-foldings. 
Note that all the higher point diagrams remain perturbatively weak during 
this entire period,
\begin{equation}
\lim_{Ht \rightarrow \lambda^{-1/2}} V^L_{2N} \sim \lambda^{N/2} \; .
\end{equation}
It should therefore be valid to use perturbation theory almost up to $\ln(\O)
= 1/\sqrt{\lambda}$.

\vskip 1cm
\noindent {\bf Note Added:} After the completion of this work we became aware 
of a highly significant paper by Starobinsky and Yokoyama \cite{SY} in which
stochastic techniques are employed to sum the leading powers of $Ht$ at each
order.

\vskip 1cm
\centerline{\bf Acknowledgments}

It is a pleasure to acknowledge stimulating and informative conversations
with L. R. Abramo. It should also be noted that we have used some of the
figure files he prepared for ref. \cite{AbWo}. This work was partially 
supported by DOE contract DE-FG02-97ER\-41029 and by the Institute for
Fundamental Theory.


\begin{thebibliography}{99}

\bibitem{Reiss} A. G. Reiss {\it et al.}, Astron. J. {\bf 116}, 1009 (1998),
{\it astro-ph/9805201}.

\bibitem{Perlmutter} S. Permutter {\it et al.}, Astrophys. J. {\bf 517}, 565 
(1999), {\it astro-ph/9812133}.

\bibitem{Hawking} S. W. Hawking and G. F. R. Ellis, The large scale structure
of space-time (Cambridge University Press, Cambridge, 1973).

\bibitem{Ford} L. H. Ford and T. A. Roman, Phys. Rev. {\bf D60}, 104018 (1999).

\bibitem{Caldwell} R. R. Caldwell, ``A Phantom Menace?'', {\it 
astro-ph/9908168}.

\bibitem{Wang} Y. Wang, Astrophys. J. {\bf 531}, 676 (2000), {\it 
astro-ph/9806185}.

\bibitem{SNAP} http://snap.lbl.gov.

\bibitem{AbWo} L. R. Abramo and R. P. Woodard, Phys. Rev. {\bf D65}, 063516 
(2002), {\it astro-ph/0109273}.

\bibitem{BD} N. D. Birrell and P. C. W. Davies, Quantum fields in curved
space (Cambridge University Press, Cambridge, 1982).

\bibitem{Peskin} M. E. Peskin and D. V. Schroeder, An Introduction to Quantum
Field Theory (Addison-Wesley, Reading, MA, 1995).

\bibitem{Schwinger} J. Schwinger, J. Math. Phys. {\bf 2}, 407 (1961).

\bibitem{Jordan} R. D. Jordan, Phys. Rev. {\bf D33}, 444 (1986).

\bibitem{BAAF} B. Allen and A. Folacci, Phys. Rev. {\bf D35}, 3771 (1987).

\bibitem{LFLP} L. H. Ford and L. Parker, Phys. Rev. {\bf D16}, 245 (1977).

\bibitem{FSW} S. A. Fulling, M. Sweeny and R. M. Wald, Commun.  Math. Phys. 
{\bf 63}, 257 (1978).

\bibitem{AVLF} A. Vilenkin and L. H. Ford, Phys. Rev. {\bf D26}, 1231 (1982).

\bibitem{L} A. D. Linde, Phys. Lett. {\bf 116B}, 335 (1982).

\bibitem{S} A. A. Starobinsky, Phys. Lett. {\bf 117B}, 175 (1982).

\bibitem{TsWo1} N. C. Tsamis and R. P. Woodard, Class. Quant. Grav. {\bf 11},
2969 (1994).

\bibitem{TsWo2} N. C. Tsamis and R. P. Woodard, Phys. Rev. {\bf D54}, 2621 
(1996), {\it hep-ph/9602317}.

\bibitem{Linde} A. D. Linde, D. A. Linde and A Mezhlumian, Phys. Rev. {\bf 
D49}, 1783 (1994).

\bibitem{SY} A. A. Starobinsky and J. Yokoyama, Phys. Rev. {\bf D50}, 6357
(1994).

\end{thebibliography}
\end{document}